\documentclass[12pt]{article}
\usepackage{latexsym}
\usepackage{amsmath,amsbsy,amssymb}
\usepackage{verbatim}
\usepackage{bm}
\usepackage{cite}
\usepackage{ragged2e}
\usepackage{dutchcal}

\usepackage{curves}
\usepackage{epic}
\usepackage{eepic}
\usepackage{epsfig}
\newcommand{\dd}{\mbox{\rm d}}

\newcommand{\dg}{\dagger}

\newcommand{\tl}{\tilde}

\newcommand{\om}{\omega}

\newcommand{\DD}{\mbox{\rm D}}

\newcommand{\p}{\partial}

\newcommand{\be}{\begin{equation}}
\newcommand{\bear}{\begin{eqnarray}}
\newcommand{\ear}{\end{eqnarray}}
\newcommand{\ee}{\end{equation}}

\newcommand{\lbl}{\label}
\newcommand{\bi}{\bibitem}
\newcommand{\ci}{\cite}

\newcommand{\vs}{\vspace}
\newcommand{\hs}{\hspace}

\newcommand{\vphi}{\varphi}
\newcommand{\bp}{{\bm p}}

\newcommand{\bx}{{\bm x}}
\newcommand{\bX}{{\bm X}}

\newcommand{\vac}{|0 \rangle}

\newcommand{\sg}{\sigma}

\begin{document}

\begin{center}

\ 

\vs{1cm}

\baselineskip .7cm

{\bf \Large A Novel Approach to Quantum Gravity in the Presence of Matter without the Problem of Time} 

\vs{7mm}

\baselineskip .5cm
Matej Pav\v si\v c

Jo\v zef Stefan Institute, Jamova 39,
1000 Ljubljana, Slovenia

e-mail: matej.pavsic@ijs.si

\vs{9mm}

{\bf Abstract}
\end{center}

An approach to the quantization of gravity in the presence matter is examined which starts
from the classical Einstein-Hilbert action and  matter approximated by ``point'' particles minimally
coupled to the metric. Upon quantization, the Hamilton constraint assumes the form of
the Schr\"odinger equation: it contains the usual Wheeler-DeWitt term and the term with
the time derivative of the wave function. In addition, the wave function also satisfies
the Klein-Gordon equation, which arises as the quantum counterpart of the constraint
among particles' momenta. Comparison of the novel approach with the usual one in which
matter is represented by scalar fields is performed, and shown that those approaches
do not exclude, but complement each other. In final discussion it is pointed out that
the classical matter could consist of superparticles or spinning particles, described by
the commuting and anticommuting Grassmann coordinates, in which case spinor fields
would occur after quantization.

\baselineskip .45cm

{\footnotesize 
.}

\baselineskip .55cm
Keywords: Canonical quantum gravity; problem of time; wave function of the universe; quantum field theory; configuration space

\section{Introduction}

Quantum gravity is an unfinished project. 
In one actively investigated approach, the state of
the universe is represented by a wave functional that satisfies the Wheeler-DeWitt equation and comprises
matter degrees of freedom as well as the gravitational field. Matter degrees of freedom
are typically given in terms of scalar or spinor
fields\,(see, e.g., Ref.\,\ci{MatterScalarField}). However, when we
consider the evolution of the universe, we usually talk about {\it positions} of objects, e.g., we say
that galaxies are receding from us, that there is a black hole in the center of a galaxy, etc.
The works such as those considered, e.g.,
in Refs.\,\ci{BarbourOthers1,BarbourOthers2,BarbourOthers3,BarbourOthers4,BarbourOthers5,Barbour},
inspired by the implications
of the Wheeler-DeWitt equation, consider models in which the wave function of the universe is
given in terms of positions of particles. Here we provide an explicit demonstration for the first time
of how a wave functional of,
e.g., a  {\it scalar field} and a gravitational field is related to a wave functional  of
{\it many particle positions} and a gravitational field.
In the literature on quantum gravity the space of configurations of a scalar or whatever field is
called configuration space, but the same name `configuration space' is used in other branches
of physics for the configuration space of one, two three, or many particle positions.

First we analyze the conventional field theory of a scalar field in the
Schr\"odinger functional representation in which quantum fields are $c$-numbers, whilst the
corresponding canonically conjugated variables are functional derivatives. A generic state,
represented as a functional of the scalar field, can be expanded over the basis of the Fock space.
In distinction to the usual approaches, in which a generic state as a superposition
of multiparticle states in momentum space, we consider multiparticle states in position space.
The scalar field QFT state functional can then be  
generalized to include a fixed (background) metric field $q_{ij} (\bx)$ on a space like 3D hypersurface
$\Sigma$ in spacetime.

Next, the theory is formulated within a more general framework in which an unfixed, dynamical 
4D metric is taken into account and is split according to the ADM prescription. The classical action
contains a matter part, $I_{\rm m}$, plus the gravitational part $I_{\rm G}[g_{\mu \nu}]$.
For the matter part it is customary to  take a functional of some fields, such as a scalar,
spinor or Maxwell field, etc. In the case of a scalar field, $\vphi$, the action is thus
$I_{\rm m} [\vphi,g]$.

In this paper we explore an alternative procedure in which we start 
directly from a classical action $I_{\rm m} [X^\mu, g_{\mu \nu}]$ for a system of ``point particles''
coupled to a gravitational field. We point out that such a procedure makes sense, if the coordinates
$X^\mu$ are not assumed to be associated with true point particles, but with effective positions
of extended particles. In the Gupta-Bleuler quantization
the classical constraints, obtained by variation of the Lagrange multipliers, become
operator constraints on the quantum states. As usual, the Lagrange multipliers in $I_{\rm m} [\vphi,g]$ are
the lapse and shift functions $N$, $N_i$, $i=1,2,3$, whereas in $I_{\rm m} [X^\mu,g]$
we also have the Lagrange multiplier $\lambda$, associated with the $\tau$-reparametrization
invariance, which gives the mass shell constraint, and, after quantization, the Klein-Gordon
equation.

In the procedure, discussed in this paper,  in which we start from the classical action
$I_{\rm m} [X^\mu,g] + I_{\rm G}[g_{\mu \nu}]$, upon quantization the time $X^0$ does
not disappear from the equations. On the contrary, in the usual procedure that starts from the action
$I_{\rm m}[\vphi,g]+I_{\rm G}[g_{\mu \nu}$], there is manifestly no time in the quantum equations,
which is the notorious ``problem of time". In  Ref.\ci{Anderson} are reviewed many different
approaches to the resolution of this tough problem.

The paper is organized as follows. In Sec.\,2 we first briefly review some basic facts about the
Schr\"odinger functional representation for a scalar field and show how a fixed gravitational field
can be included into the description. In Sec.\,3 we consider the dynamical gravity as well, first with
the novel and then with the traditional approach to the matter term, and show how those
distinct procedures are related to each other. In Sec.\,4  we resume our findings, and discuss a
larger framework into which they can be embedded. In particular, we point out, that the classical
matter, consisting of particles, can be extended to superparticles or spinning particles that are described
not only by the commuting coordinates $X^\mu$, but also by anticommuting coordinates, in the
literature often denoted as $\xi^\mu$ or $\theta^\mu$. Coupling of such a system to gravity and
quantizing it along the lines discussed in this paper, would then bring spinor fields into the
description.

\section{Wave function in quantum field theory}

To make our presentation self-consistent we will first review some known facts from quantum field theory,
and then show a novel way of introducing the gravitational field into the description of quantum states.
But first let us recall how in flat spacetime we arrive at a quantum field theory from a relativistic point particle. It is described by a ``minimal'' length action that can be cast into the following equivalent form:
\be
  I[X^\mu (\tau) = \frac{1}{2} \int \dd \tau \left ( \frac{{\dot X}^\mu{\dot X}\mu} {\lambda}+ \lambda  m^2 \right ),
\lbl{2.1}
\ee
where $\lambda$ is a Lagrange multiplier, variation of which gives the constraint $\left (p^\mu p_\mu - m^2 \right ) = 0$.
Upon quantization it becomes the Klein-Gordon equation  
\be
  \left (\p_\mu \p^\mu + m^2) \vphi (x) \right )= 0.
\lbl{2.1a}
\ee
The latter can be derived from the action for a scalar field,
\be
  I[[\vphi(x)] = \frac{1}{2} \int \dd^4 x \left ( \p_\mu \vphi \p^\mu \vphi  - m^2 \vphi^2.
  \right ),
\lbl{2.1b}
\ee
Here $x\equiv x^\mu = (t,\bx)$, $\mu=0,1,2,3$, $\bx \equiv x^i$, $i=1,2,3$, 
are spacetime coordinates.
The canonically conjugated variables, $\vphi(t,\bx)$ and
$\Pi(t,\bx) = {\dot \vphi}(t,\bx)$, where dot denotes the derivative with respect to
the time $t$, become upon quantization the operators ${\hat \vphi}(t,\bx)$,
${\hat \Pi} (t,\bx)$, satisfying the equal time commutation relations
\be
  [{\hat \vphi}(t,\bx),{\hat \Pi} (t,\bx)] = i \delta^3 (\bx - \bx'),
\lbl{A3}
\ee
\be
  [{\hat \vphi}(t,\bx),{\hat \vphi} (t,\bx)] = 0~,~~~~~~[{\hat \Pi}(t,\bx),{\hat \Pi} (t,\bx)] = 0.
\lbl{A4}
\ee

A general solution of the Klein-Gordon equation (\ref{2.1a}) can  be expanded according
to\footnote{We use the normalization as adopted, e.g., in the textbook by Peskin\,\ci{Peskin}}
\be
  {\hat \vphi}(t,\bx) = \int \frac{\dd^3 \bp}{\sqrt{(2 \pi)^3 2 \om_\bp}}
  \left ( a(\bp) {\rm e}^{- i p x} + a^\dg (\bp) {\rm e}^{i p x} \right ) ,
\lbl{A4a}
\ee
where $\om_\bp = \sqrt{m^2 + \bp^2}$ and $px \equiv k_\mu x^\mu$. The operators
$a(\bp)$, $a^\dg (\bp)$ are consistent with (\ref{A3}), (\ref{A4}) if they satisfy
\be
  [a (\bp),a^\dg (\bp')] = \delta^3 (\bp-\bp') ,
\lbl{A5}
\ee
\be
  [a (\bp),a (\bp')] = 0 ~,~~~~~~~ [a^\dg (\bp),a^\dg (\bp')] = 0 .
\lbl{A5a}
\ee

In the $\varphi (x)$ representation  the field operator is represented by a $c$-number,
time independent, field $\vphi(\bx)$, whereas the conjugate momentum operator
is represented by the functional derivative\,\ci{Jackiw,Hatfield}:
\be
  {\hat \vphi}(\bx) \longrightarrow \vphi (\bx)~,~~~~~{\hat \Pi} (\bx) \longrightarrow
  - i \frac{\delta}{\delta \vphi (\bx)}.
\lbl{A7}
\ee

A state $|\Psi \rangle$ is represented by a time dependent wave functional
$\Psi[t,\vphi(\bx)] \equiv \langle \vphi(\bx)|\Psi \rangle$, which satisfies
the Schr\"odinger functional equation
\be
  i \frac{\p \Psi [t,\vphi(\bx)]}{\p t} = H \Psi [t,\vphi(\bx)] .
\lbl{A6}
\ee
The Hamilton operator is (see e.g.,Ref.\,\ci{Hatfield})
\be
  H = \int \dd^3 \bx \left ( {\hat \Pi} {\dot {\hat \vphi}} - {\cal L} \right ) = 
  \frac{1}{2} \int \dd^3 \bx \left ( - \frac{\delta^2}{\delta \vphi(\bx)^2} 
  + \vphi (m^2 - \nabla^2 )\vphi \right )
\lbl{A7a}
\ee
One could object that the second order functional derivatives give singularity. And yet, this is
precisely what happens in quantum field theory in whatever representation. Namely,
stacionary solution of Eq.\,(\ref{A7a}) satisfy
\be
  H \Psi[\vphi(\bx) ]= E \Psi[\vphi(\bx)] .
\lbl{A7b}
\ee
An example of a solution is the vacuum functional
\be
  \Psi_0[\vphi(\bx) = \eta \, {\rm exp} \left [- \frac{1}{2} \int \dd^3 \bx \vphi(\bx) 
  \sqrt{m^2 - \nabla^2} \vphi (\bx)  \right ]  ,
\lbl{A7c}
\ee
which, when inserted into the stacionary Schr\"odinger equation (\ref{A7b}), gives
\be
  \frac{1}{2} \int \dd^3 \bx \left ( \sqrt{m^2 - \nabla^2} \, \delta (0) \right ) \Psi_0 = E \Psi_0 .
\lbl{A7d}
\ee
This corresponds to the singular (``zero point'') energy of the vacuum.

However, in Ref.\,\ci{PavsicBrane1,PavsicBrane2} (see also\,\ci{PavsicBook})
the following generalization of the Hamilton operator (\ref{A7a}) was considered:
\be
  H= \frac{1}{2} \left ( - \rho^{a(\bx) b (\bx')} \p_{a(\bx)} \p_{b(\bx')} 
  + \vphi^{a(\bx)} \om_{a(\bx) b (\bx')}  \vphi^{b(\bx')} \right ) ,
\lbl{A7e}
\ee
where $\rho^{a(\bx) b (\bx')}$ is a metric in the space of fields $\vphi^{a(\bx)} \equiv \vphi^a (\bx)$.
Then, in general, no singularity arises, despite that the expression contains the second order
functional derivatives $\p_{a(\bx)} \equiv \frac{\p}{\p \vphi^{a(\bx)}}$. As to the ordering ambiguity, one can
extend the procedure of Ref.\,\ci{PavsicOrder} by introducing the basis vectors $h_{a(\bx)}$,
satisfying the Clifford algebra relations
\be
  h_{a(\bx)} \cdot h_{b(\bx')} \equiv \frac{1}{2} \left (  h_{a(\bx)} h_{b(\bx')}+  h_{b(\bx')} h_{a(\bx)} \right )
  = \rho_{a(\bx) b(\bx)},
\lbl{A7f}
\ee
and, instead of (\ref{A7e}), define the following Hamilton operator, covariant in function space:
\be
  H= -\frac{1}{2} \left (  \left ( h^{a(\bx)} \p_{a(\bx)} \right ) \left ( h^{b(\bx')} \p_{b(\bx')} \right )
  +\vphi^{a(\bx)} \om_{a(\bx) b (\bx')}  \vphi^{b(\bx')} \right ).
\lbl{A7g}
\ee
Here $\om_{a(\bx) b (\bx')}$ is a coupling between the fields that generalizes $\sqrt{m^2 - \nabla^2}$,
The momentum vector operator $\Pi = h^{a(\bx)} \Pi_{a(\bx)} = - i h^{a(\bx)} \p_{a(\bx)}$ is then
Hermitian\,\ci{PavsicOrder}.

We are not interested here into the mathematical intricaces\,\ci{Engle} concerning hermiticity
of the momentum operator ${\hat \Pi} (\bx)$ that must be taken into account in the cases of generic
function spaces. For us it is important that the definition (\ref{A7}) works well in the
relevant physical calculations as nicely shown in Ref.\,\ci{Hatfield}.
The expectation value is
\be
  \langle {\hat \Pi} (\bx) \rangle = \int {\cal D} \vphi (\bx) \Psi^* [\vphi (\bx)] (-i) \frac{\delta}{\delta \vphi (\bx)}
  \Psi[\vphi (\bx)],
\lbl{BB1}
\ee
where for the measure we take
\be
  {\cal D} \vphi (\bx) = \prod_\bx \dd \vphi (\bx) ,
\lbl{BB2}
\ee
which holds in the function space with the metric
\be
    \rho (\bx,\bx') = \delta^3 (\bx - \bx'),
\lbl{BB3}
\ee
so that the squared distance element in the function space is
\be
   \int \dd^3 \bx \dd^3 \bx'  \rho (\bx,\bx') \dd \vphi (\bx)  \dd \vphi (\bx')
   =\int \dd^3 \bx \, \dd \vphi (\bx)^2 .
\lbl{BB4}
\ee
Taking the complex conjugate, we have
 $$ \langle {\hat \Pi} (\bx) \rangle^* = \int {\cal D} \vphi (\bx) \Psi [\vphi (\bx)] i \frac{\delta}{\delta \vphi (\bx)}
  \Psi^*[\vphi (\bx)]  \hs{2cm}$$
\be
   \hs{3cm} = \int {\cal D} \vphi (\bx) \left [\Psi^* [\vphi (\bx)] (-i) \frac{\delta}{\delta \vphi (\bx)}\Psi[\vphi (\bx)]
   - i \frac{\delta}{\delta \vphi (\bx)} (\Psi^* \Psi) \right ]  
\lbl{BB5}
\ee
Let us assume that  $\Psi^*[\vphi(\bx)] \Psi [\vphi(\bx)] \to 0$ if $\vphi \to \infty$, which means that the state is
localized around a fixed, expected field configuratiion
Then the expectation value is real and the momentum operator is Hermitian. If we also
assume that $\int {\cal  D} \vphi(\bx)\,\Psi^*[\vphi(\bx)] \Psi [\vphi(\bx)] = 1$, then $\Psi[\vphi(\bx)]$ can be
interpreted as the probability amplitude, and $\Psi^*[\vphi(\bx)] \Psi [\vphi(\bx)]$ as the probability density
for the field configuration $\vphi (\bx)$, whose expectation value is
\be
  \langle \vphi(x) \rangle 
  = \int {\cal D} \vphi (\bx) \Psi^* [\vphi (\bx)] \vphi (x)  \Psi[\vphi (\bx)] .
\lbl{BB6}
\ee

Using (\ref{A7}), the operators $a(\bp)$, $a^\dg (\bp)$ can be represented as follows\,\ci{Jackiw}:
\be
  {a}^\dg (\bp) =  \int \dd^3 \bx \, {\rm e}^{-i \bp \bx} \left ( \sqrt {(2 \pi)^3 2 \om_\bp } \vphi (\bx) -
  \frac{1}{\sqrt {(2 \pi)^3 2 \om_\bp }} \frac{\delta}{\delta \vphi (\bx)} \right ),
\lbl{4.11}
\ee
\be
 {a}(\bp) =  \int \dd^3 \bx \, {\rm e}^{i \bp \bx} \left ( \sqrt {(2 \pi)^3 2 \om_\bp } \vphi (\bx) +
  \frac{1}{\sqrt {(2 \pi)^3 2 \om_\bp }} \frac{\delta}{\delta \vphi (\bx)} \right ),
\lbl{4.12}
\ee

If we take the Fourier transform of the  operators $a^\dg(\bp)$, $a(\bp)$,
namely
\be
  {a}^\dg (\bx) = \frac{1}{\sqrt{(2 \pi)^3}}\int \dd^3 \bp \, {\rm e}^{i \bp \bx} {a}^\dg (\bp) 
\lbl{4.15}
\ee
\be
  {a}(\bx) = \frac{1}{\sqrt{(2 \pi)^3}}\int \dd^3 \bp \, {\rm e}^{-i \bp \bx} {a}(\bp) 
\lbl{4.16}
\ee
we obtain
\be
  [{a}(\bx),{a}^\dg (\bx')] =  \delta^3 (\bx-\bx'),
\lbl{4.17}
\ee
\be
  [{a}(\bx),{ a}(\bx')]=0~,~~~~~~ [{a}^\dg (\bx),{a}^\dg (\bx')]=0.
\lbl{4.18}
\ee
These are the commutation relations for creation (annihilation) operators that
create (annihilate) a particle at $\bx$. From (\ref{4.17}),(\ref{4.18}), by using 
(\ref{4.11}),(\ref{4.12}), we find the following explicit expressions:
\be
  a^\dg (\bx) = \sqrt{ (2 \pi)^3 \,2} (m^2-\nabla^2)^{1/4} \vphi(\bx) 
  - \frac{1}{ \sqrt{(2 \pi)^3 \, 2} (m^2-\nabla^2)^{1/4}}\,\frac{\delta}{\delta \vphi (\bx)}
\lbl{4.18a}
\ee
\be
  a (\bx) = \sqrt{(2 \pi)^3\,2} (m^2-\nabla^2)^{1/4} \vphi(\bx) 
  +\frac{1}{ \sqrt{(2 \pi)^3 \,2} (m^2-\nabla^2)^{1/4}}\,\frac{\delta}{\delta \vphi (\bx)}
\lbl{4.18b}
\ee

From the latter equations we obtain
\be
  \vphi (\bx) = \frac{1}{ \sqrt{(2 \pi)^3 \,2} (m^2-\nabla^2)^{1/4}}\,(a (\bx) + a^\dg (\bx)),
\lbl{4.18c}
\ee
\be
  \frac{\delta}{\delta \vphi (\bx)} = \sqrt{(2 \pi)^3 \,2} (m^2-\nabla^2)^{1/4}(a (\bx) - a^\dg (\bx)) .
\lbl{4.18d}
\ee
Using (\ref{4.18c}),(\ref{4.18d}), we can express the Hamiltonian (\ref{A7a}) in terms
of $a(\bx)$ and $a^\dg (\bx)$:
\be
  H = \frac{1}{2} \int \dd^3 \bx \,  \left ( a^\dg (\bx) \sqrt{m^2 - \nabla^2} \, a(\bx)
  +   \left ( \sqrt{m^2 - \nabla^2} a (\bx) \right ) a^\dg (\bx) \right ) .
\lbl{4.18e}
\ee
If expressed in terms of $a(\bp)$, $a^\dg (\bp)$, the Hamiltonian has the usual form,
\be
 H = \frac{1}{2} \int \dd^3 \bp \, \om_\bp \left ( a^\dg (\bp) a (\bp) + a(\bp) a^\dg (\bp) \right ).
\lbl{4.18f}
\ee

A generic state is a superposition of many particle states:
\be
  \Psi[t, \vphi (\bx)] = \sum_{r=0}^\infty \int \dd^3 \bp_1 \dd^3 \bp_2 ...\dd^3 \bp_r
 {\phi} (t, \bp_1,\bp_2,...,\bp_r) a^\dg (\bp_1) a^\dg (\bp_2) ... a^\dg (\bp_r) \Psi_0 [\vphi(\bp)],
\lbl{4.19a}
\ee
where ${\phi} (t,\bp_1,\bp_2,...,\bp_r)$ is a complex valued wave packet profile for
an $r$-particle state in momentum representation

In terms of the Fourier transformed creation operators (\ref{4.15}),(\ref{4.16}),
the same state reads:
\be
  \Psi[t,\vphi (\bx)] = \sum_{r=0}^\infty \int \dd^3 \bx_1 \dd^3 \bx_2 ...\dd^3 \bx_r
  \psi(t, \bx_1,\bx_2,...,\bx_r) a^\dg (\bx_1) a^\dg (\bx_2) ... a^\dg (\bx_r) \Psi_0 [\vphi(\bx)],
\lbl{4.19}
\ee
where $\psi(t,\bx_1,\bx_2,...,\bx_r)$ is a complex valued wave packet profile for
an $r$-particle state in position representation. Its absolute square $|\psi|^2 = \psi^* \psi$
gives the probability density of observing at a time $t$ a multi particle configuration at positions
$\bx_1,\bx_2,...,\bx_r$, and as discussed in the next paragraph and the references cited therein,
the apparent non covariance of $|\psi (\bx)|^2$ is not problematic.

Within relativistic quantum field theories, the wave packet profiles in momentum
representation are occasionally used in the literature  when considering the $S$-matrix.
In the textbook by Peskin and Schr\"oder\,\ci{Peskin}, a single particle wave packet
is considered, and it is mentioned that $\phi (\bp)$ is the Fourier transform of the spatial
wavefunction. In general, a wave packet depends on time.
In a given Lorentz reference frame one can thus consider either $\phi (t,\bp)$
or $\psi (t,\bx)$ (and their multiparticle extensions), the corresponding creation operators
being $a^\dg (\bp)$ or $a^\dg (\bx)$. Although asymptotic states are usually taken to have
definite momenta\footnote{
A state with definite momentum, $a^\dg (\bp) \vac \equiv |\bp \rangle$,
is an approximation. In reality, there is always some spreading. A sharp momentum state is not
in Hilbert space, it belongs to generalized states.},
   there are situations, when they form wave packet profiles, in which case
it is possible to measure the detection time, the time of flight, etc. Such wavepacket profiles,
of course, can be observed from different Lorentz reference frames. When observed from
another Lorentz frame, the wave packet, and hence the probability density, looks different,
it is subjected to an appropriate transformation, but it is still spread
around a centroid momentum, or equivalently, around a centroid position\,\ci{PavsicLocal}.

The vacuum state satisfies
\be
  a(\bx) \Psi_0 [\vphi(\bx)] = 0.
\lbl{4.20}
\ee
Using Eq.\,(\ref{4.18b}) in the latter equation, we find for the solution
the vacuum functional (\ref{A7c}) that was obtained directly from the stacionary
Schr\"odinger equation (\ref{A7b}).

Usually authors do not work in the coordinate ($\bx$), but in the momentum
($\bp$) representation, which in many respects is more practical. In particular,
in momentum representation it is straightforward to calculate the normalization
constant for the vacuum functional. But for the purposes of our paper,
the coordinate representation is useful, because once the expressions
(\ref{4.18a}),(\ref{4.18b}) and (\ref{A7c}) are obtained, we can generalize them to include
the metric field $q_{ij} (\bx)$ on a 3D hypersurface $\Sigma$ by replacing $\nabla^2$ with the
covariant operator $q^{ij} D_i D_j$, which when acting on a scalar field gives
$ \frac{1}{\sqrt{q}} \p_i (\sqrt{q} \p^i \vphi (\bx))$, where $q={\rm det}\, q_{ij}$,
i.e., the determinant of the 3-metric. To our knowledge such procedure has not yet
been considered in the literature, and in the following we will show its usefulness in
introducing a gravitational field into the description.

If we make the replacement
\be
    \nabla^2 \rightarrow q^{ij} D_i D_j = D_i D^i \equiv D^2 ,
\lbl{4.21}
\ee
then the creation and annihilation operators (\ref{4.18a}),(\ref{4.18b}) become
functionals of $q_{ij}(\bx)$, and so does the vacuum functional (\ref{A7c}):
\be
  a(\bx) \rightarrow a[\bx,q_{ij}(\bx)]  , \lbl{4.q2}\ee
\be  
  a^\dg(\bx) \rightarrow a^\dg[\bx,q_{ij}(\bx)]  ,\lbl{4.q3}\ee  
\be
 ~~ \Psi_0[\vphi (\bx)] \rightarrow \Psi_0 [\vphi (\bx), q_{ij}(\bx)] . \lbl{4.q4}
\ee
It turns out that they satisfy the same commutation relation (\ref{4.17}),(\ref{4.18})

The so modified operator $a^\dg[\bx,q_{ij}(\bx)]$ creates a particle\footnote{
It is usually stated that the notion of particle depends on the metric, and by ``particle'' it is
understood an excitation with definite momentum. Here by ``particles'' we mean just the excitations, created
or annihilated by the modified position dependent operators (\ref{4.q2}),(\ref{4.q3}).}
at position $\bx$ in the
gravitational field $q_{ij}(\bx)$. One finds that the modified operator
$a[\bx,q_{ij}(\bx)]$ annihilates the modified vacuum functional  $\Psi_0 [\vphi(\bx), q_{ij}(\bx)]$.
A generic state can then be represented by the following functional:
\bear
  &&\Psi[t,\vphi(t,\bx),q_{ij} (\bx)] = \sum_{r=1}^{\cal \infty} \int \dd^3 \bx_1 ...
\dd^3 \bx_r   \psi[t,\bx_1,...,\bx_r, q_{ij} (\bx)] \nonumber\\
  &&\hs{4cm}\times a^\dg [\bx_1,q_{ij} (\bx)]...a^\dg [\bx_r,q_{ij}(\bx)] \Psi_0 [\vphi (\bx),q_{ij} (\bx)] .
\lbl{4.q5}
\ear
Here $\psi[t,\bx_1,...,\bx_r, q_{ij} (\bx)]$ is the amplitude for the probability
that  we will find matter particles at positions $\bx_1,\bx_2,...,\bx_r$
in a gravitational field $q_{ij}(\bx)$. This is thus a wave functional
that depends on particle positions and on the corresponding gravitational  field at
those positions. In the case when only one particle is created, its wave functional
is $\psi[\bx_1, q_{ij} (\bx)]$. The state (\ref{4.q5}) is then given by
\be
 \Psi[t,\vphi(\bx),q_{ij} (\bx)] = \int \dd^3 \bx_1
  \psi[t,\bx_1, q_{ij} (\bx)] a^\dg [\bx_1,q_{ij}(\bx)] \Psi_0 [\vphi (\bx),q_{ij} (\bx)] .
\lbl{4.q6}
\ee
In Ref.\,\ci{PavsicKleinWheeler} the wave functional of the form $\psi[X^\mu, q_{ij} (\bx)]$,
which depends on a particle position and a gravitational field,
was considered, and found to satisfy the Klein-Gordon and the Wheeler-DeWitt
equation. This was an alternative to the usually considered state functional
$\Psi[\vphi(\bx),q_{ij} (\bx)]$ that depends on a scalar field $\vphi (\bx)$ instead on $\bx$.
In Eq.\,(\ref{4.q6}) we have a translation between those two possible
descriptions, and in Eq.\,({\ref{4.q5}) we have a translation for the general case of many
particles. In one description we have a functional $\Psi[t,\vphi(\bx),q_{ij} (\bx)]$, and in the
other description we have a set of functionals $ \psi[t,\bx_1, q_{ij} (\bx)]$,
$ \psi[t,\bx_1, \bx_2, q_{ij} (\bx)]$, ... , $ \psi[t,\bx_1,...,\bx_{\cal N}, q_{ij} (\bx)]$,
which are components of the expansion of the state functional in terms of the
basis states $ a^\dg [\bx_1,q_{ij} (\bx)]...a^\dg [\bx_r,q_{ij}(\bx)] \Psi_0 [\vphi (\bx),q_{ij} (\bx)]$,
$r=0,1,2,...,{\cal N}$, where ${\cal N}$ is arbitrary and can go to infinity.

\vs{2mm}

{\it Schr\"odinger equation}

If we substitute the expression (\ref{4.19}) for the state functional and (\ref{4.18e}) for the
Hamiltonian into the Schr\"odinger equation (\ref{A6}), we obtain a set of equations
for the multiparticle wavefunctions:
\be
  i \frac{\p \psi ((t,\bx_1,...,\bx_r)}{\p t} = \sum_{\ell=1}^r \sqrt{m^2-\nabla_{\bx_\ell}^2}\, \psi \, (t,\bx_1,...,\bx_r)~,
  ~~~r=1,2,...,\infty,
\lbl{B1a}
\ee
where
\be
  \nabla_\ell^2 \equiv - \frac{\p^2}{\p x_\ell^i \p x_{i \ell}}~,~~~~i = 1,2,3,
\lbl{B2a}
\ee
In Eq.\,(\ref{B1a}) we have omitted the infinite zero point energy, because it cancels out
in the expressions containing the probability density $\psi^* \psi$.

If we take into account also the gravitational field $q_{ij} (\bx)$, the Schr\"odinger equation
(\ref{A6}) generalizes so that instead of $\Psi[t,\vphi(\bx)]$ we have $\Psi[t,\vphi(\bx),q_{ij}(\bx)]$
and in the Hamiltonian (\ref{4.18e}) the operator $\nabla^2$ generalizes according to (\ref{4.21}):
\be
  i \frac{\p \Psi [t,\vphi (\bx),q_{ij} (\bx)]}{\p t} = H  \Psi [t,\vphi (\bx),q_{ij} (\bx)] .
\lbl{B3a}
\ee
Then, instead of (\ref{B1a}) we obtain
\be
  i \frac{\p \psi [t,\bx_1,...,\bx_r,q_{ij}(\bx)]}{\p t}
   = \sum_{\ell=1}^r \sqrt{m^2-\DD_{{\bx}_\ell}^2} \, \psi \, [t,\bx_1,...,\bx_r,q_{ij}(\bx)],~~ ~r=1,2,...,\infty.
\lbl{B4a}
\ee

Equation (\ref{B3a}), or equivalently, (\ref{B4a}), describes the evolution of a many particle state in
the presence of a bakground gravitational field $q_{ij}(\bx)$. At a certain time $t$, the probability density
$|\psi \, [t,\bx_1,...,\bx_r,q_{ij}(\bx)]|^2$ is centered around a configuration
$\bx_1,...,\bx_r$ and a 3-metric $q_{ij}(\bx)$. At another time $t$, it is centered around a
different configuration and a different intrinsic metric. In other words,
$\int_\Omega |\psi \, [t,\bx_1,...,\bx_r,q_{ij}(\bx)]|^2 \dd^3 \bx_1 \dd^3 \bx_2 ... \dd^3 \bx_r\, {\cal D} q_{ij}(\bx)$
is the probability that at time $t$  we find $r$ particles within a domain $\Omega$ around
the positions $\bx_1,\bx_2,...,\bx_r$ and in the gravitational field $q_{ij}(\bx)$.
Because in the right hand side of Eq.\,(\ref{B4a}) there is no operator term acting on $q_{ij} (\bx)$,
the probability density remains at all times $t$ centered around the same 3-metric $q_{ij}(\bx)$.
In particular, it can be $q_{ij} (\bx) = \delta_{ij}$, but in general it is a position dependent metric
with a non vanishing 3-curvature. Equation (\ref{B4a}) then describes evolution of a wave
function in a fixed non trivial gravitational field. There is no dynamics of $q_{ij}(\bx)$ itself in such a formalism

\section{Dynamics of matter coupled to gravity}

\subsection{A novel approach}

In the previous section we started from a relativistic point particle in flat spacetime
and arrived upon first quantization at the Klein-Gordon
equation for a scalar field $\vphi (x)$. Then we quantized the field $\vphi (x)$ as well,
and arrived at quantum field theory in which a generic state can be represented as a functional
$\Psi[t,\vphi (\bx)]$, expanded in terms of many particle states with wave packet profiles
(wave functions) $\phi(t,\bp_1,\bp_2,...,\bp_r)$, or equivalently,  $\psi(t,\bx_1,\bx_2,...,\bx_r)$,
satisfying the Schr\"odinger equation (\ref {B1a}). We then discussed how such
wavefunctions can be generalized to include a fixed gravitational field. We have thus arrived
at the quantum evolution of a many particle wave function (\ref{B4a}) in a fixed 3-metric field $q_{ij}(\bx)$
on a simultaneity hypersurface $\Sigma$. In order to include a dynamics of the gravitational
field $g_{\mu \nu} (x)$, $x \equiv x^\mu$, $\mu=0,1,2,3$, and consequently of
the induced metric $q_{ij}(\bx)$ on $\Sigma$, let us consider the classical
action for a many particle system coupled to gravity:
\be
  I[X_n^\mu,g_{\mu \nu} (x),\lambda_n] = I_{\rm m} [X_n^\mu,g_{\mu \nu} (x),\lambda_n]+ I_{\rm G} [g_{\mu \nu}(x)] ,
\lbl{3.1a}
\ee
where\footnote{The parameter $\tau$ is arbitrary  and in principle different on each worldline.
To simplify the notation we write $\tau$ instead of $\tau_n$,}
\be
  I_{\rm m} [X_n^\mu,g_{\mu \nu} (x),\lambda_n] = \frac{1}{2} \sum_{n=1}^{\cal N} \int \dd \tau
  \left ( \frac{{\dot X}_n^\mu {\dot X}_n^\nu g_{\mu \nu}}{\lambda_n} + \lambda_n m_n^2 \right ) ,
\lbl{3.1b}
\ee
and
\be
  I_{\rm G} = \kappa \int \dd^4 x \sqrt{-g}\, R ,~~~~~~~ \kappa = (16 \pi G)^{-1}.
\lbl{3.1c}
\ee
Here the functions $X_n^\mu (\tau)$ of an arbitrary monotonically increasing parameter
$\tau$ describe the worldlines associated with positions of particles, e.\,g., their centers of mass.
Realistic particles are not point like, they are extended beyond their
Schwarzschild radii\footnote{
The action (\ref{3.1b}) can also describe a system of (mini) black holes with their positions
being parametrized with $X_n^\mu$ and tracing the worldlines $X_n^\mu (\tau)$. Such a system,
instead of being considered within a complicated detailed description involving the mutual
dynamics of black holes' gravitational field, could be as well approximately described in terms
of their positions $X_n^\mu$. The very fact that we talk about a black hole in the center of our
galaxy, or in the center of another galaxy, means that we ascribe to a black hole a position, and
parametrize it by a set of coordinates.},
but in our description we take into account only the particle's center of mass.

Despite that the definition of the center of mass is a subject of controversy\ci{Price}, for our
purpose here it is important that each of the objects whose motion is governed by the action (\ref{3.1b}) is
extended, not point like, and that can be described by four coordinates only, so that the infinitely many
degrees of freedom of an extended object are neglected.
How this works for an extended object confined within a narrow tube in spacetime
is shown in the derivation of the Papapetrou equation\,\ci{Papapetrou}. The derivation is based on
the moments of the stress-energy tensor around a chosen worldline within the tube. If only the
first moment (``monopole'') is taken into account, one obtains the geodesic equation for such a worldline.
In such sense one should also consider the wordlines occurring in the action (\ref{3.1b}). For further support of
our argument, see Appendix A.

In the continuum limit of many densely packed worldlines the action (\ref{3.1b}) becomes the dust action.
Coupling of dust to gravity was considered by Brown and Kuchar\,\ci{Brown} in order to resolve the problem of
time in quantum gravity. In their approach dust is supposed to be present everywhere and its degrees of freedom
incorporate time.  We will show that instead of dust one can as well employ a system of ``point'' particles that even need not be present everywhere. It comes out that the generator of time translations is directly
associated with the stress-energy tensor of such system of particles. 

If one performs the ADM split of spacetime, then the action (\ref{3.1a}) can be written as a
functional of the 3-metric $q_{ij}$, and the lapse and shift functions, $N$ and $N^i$, $i,j=1,2,3$.
Rewriting the matter part of the action, (\ref{3.1b}), into the phase space form,
\be
  I_{\rm m} [X_n^\mu,p_{n \mu},\lambda_n, g_{\mu \nu}]
  = \sum_n \int \dd \tau \left ( p_{n \mu} {\dot X}_n^\mu
      - \frac{\lambda_n}{2} (g_{\mu \nu} p_n^\mu p_n^\nu - m_n^2 ) \right ),
\lbl{3.2}
\ee
and performing the ADM split, we obtain\,\ci{PavsicKleinWheeler}
\bear
   &&I_{\rm m} [X_n^\mu,p_{n \mu},\lambda_n, q_{ij},N,N^i] = 
  \sum_n \int \dd \tau \left ( p_{n \mu} {\dot X}_n^\mu \right . \hs{2cm} \nonumber \\
      &&\hs{2cm}   -  \left . \frac{\lambda_n}{2} \left [ \frac{1}{N^2} (p_{n0} - N^i p_{ni})(p_{n0} - N^j p_{nj})
     - q^{ij} p_{ni} p_{nj} -m_n^2 \right ] \right )
\lbl{3.3}
\ear
In the above action, momenta are the quantities, variation  of which gives the relation
$p_n^\mu = {\dot X}_n^\mu/\lambda_n$.

In order to have the matter action on the same footing as the gravitational action,
we must insert $1 = \int \sqrt{-g}\, \dd^4 x \,\delta^4 (x - X_n (\tau))/\sqrt{-g}$
$= \int \dd^4 x \, \delta^4 (x - X_n (\tau)) $.  But our particle is not exactly
point like, it is extended, e.g., a ball, whose worldvolume is described by $X^\mu (\tau, \sg)$,
$\sg \equiv \sg^a = (R,\vartheta,\vphi)$, $0\le R\le R_0$, $0 \le \vartheta \le \pi$,
$0 \le \vphi \le 2 \pi$, where $R_0$ is greater than the Schwarzschild radius. Therefore,
$\delta^4 \left ( x - X_n (\tau) \right )$ should be considered as an approximation to 
$\int \dd^3 \sg \sqrt{- {\bar f}} \,\delta^4 (x - X_n (\tau, \sg))$, where
${\bar f} \equiv {\rm det} {\bar f}_{ab}$, $\p_a \equiv \p/\p \sg^a$
(see also\,\ci{InfeldPlebanski}, where the so called
``good'' $\delta$-function is defined). Despite that such a $\delta$-function associated with a ball is not
{\it invariant}, it is {\it covariant}\footnote
{For instance, if we consider the case of {\it Minkowski space}, then
in another Lorentz reference frame the
$\delta$-function, corresponding to a ball-like extended object, looks different (no longer associated
with a ball at rest, but a moving ellipsoid).
However, in a new Lorentz frame one can have with respect to the new simultaneity 3-surface  a ball-like extended object,
described by $X_n (\tau, \sg)$, the corresponding delta-function being exactly
the same as our ``original'' delta-function. The concept of the  delta-function,
though not invariant, is {\it covariant}.}.
In Appendix A we consider the dynamics of a ball, modelled by
a space filling brane, described by the action that is covariant with respect to
general coordinate transformation of $x^\mu$ and $\xi^A = (\tau,\sg^a)$.

For the gravitational part of the action we obtain\,\ci{WheelerDeWitt}
\be
  I_{\rm G} [q_{ij},\pi^{ij},N,N^i] = \int \dd t\, \dd^3 \bx \left [ \pi^{ij} {\dot q}_{ij} -
  N {\cal H}_{\rm G} (q_{ij},\pi^{ij}) - N^i {\cal H}_{{\rm G} i} (q_{ij},\pi^{ij}) \right ] ,
\lbl{3.4}
\ee
where
\bear
  && {\cal H}_{\rm G} = - \frac{1}{\kappa}\, G_{ij \, k \ell} \pi^{ij} \pi^{k \ell}
  + \kappa \sqrt{q} R^{(3)} , \lbl{3.5}\\
  &&{\cal H}_{\rm G}^i = - 2 \DD_j \pi^{ij} , \lbl{3.6}
\ear
and
\be
   G_{ij\, k \ell} = \frac{1}{2 \sqrt{q}}\, (q_{ik}q_{j \ell} + q_{i \ell}q_{jk}
   -q_{ij} q_{k \ell}) .
\lbl{3.6a}
\ee
is the Wheeler-DeWitt metric.

Variation of $I=I_{\rm m} + I_{\rm G}$ with respect to $\lambda_n$, $N$, and $N^i$, which have
the role of Lagrange multipliers, gives the constraints\,\ci{PavsicKleinWheeler} that involve both
the gravitational and matter degrees of freedom:
\bear
  &&\delta \lambda_n\,:~~~~~ \frac{1}{N^2} (p_{n0} - N^{i} p_{ni})(p_{n0} - N^j p_{nj})
     - q^{ij} p_{ni} p_{nj} -m_n^2  ,  \lbl{3.8}\\
  &&\delta N\,:~~~~~ {\cal H}_{\rm G} = - \sum_n \int \dd \tau \, \lambda_n \frac{1}{N^3}
  \delta^4 (x-X_n (\tau)) (p_{n0} - N^{i} p_{ni})(p_{n0} - N^j p_{nj})
  \nonumber\\
 &&\hs{2.5cm} = - \sum_n \delta^3 (\bx - \bX_n ) \frac{1}{N} (p_{n 0} - N^i p_{n i}) \lbl{3.9} \\
  &&\delta N^i\,: ~~~~~ {\cal H}_{{\rm G}i} = - \sum_n \int \dd \tau \, \lambda_n \, \frac{1}{N^2}
   \delta^4 (x - X_n (\tau)) p_{ni} (p_{n0} - N^j p_{nj}), \nonumber \\
  &&\hs{2.7cm}= - \sum_n \delta^3 (\bx - \bX_n) p_{ni} \lbl{3.10}
\ear
In Eq.\,(\ref{3.9}) we used $p_n^0 = g^{0 \nu}p_{n \nu} = (1/N^2) (p_{n 0} - N^i p_{ni})$, replaced
 $(p_n^0)^2$ with $ {\dot X}_n^0/\lambda_n$, and then proceeded as follows:
$$
  \int \dd \tau \,  \delta (x^0-X_n^0 (\tau))
  \delta^3 (\bx - \bX_n (\tau))  \frac{1}{N} (p_{n0} - N^{i} p_{ni}){\dot X}_n^0 $$
\be
  = \int \dd \tau\,  \frac{1}{|{\dot X}_n^0|} \delta^3 (\bx - \bX_n (\tau)) \frac{1}{N}
  (p_{n0} - N^{i} p_{ni}) {\dot X}_n^0 \Big|_{\tau=\tau_c}
  = \delta^3 (\bx - \bX_n ) \frac{1}{N} (p_{n 0} - N^i p_{n i}),
\lbl{3.9a}
\ee
where $\tau_c$ is the solution of the equation $x^0 = X_n^0 (\tau)$, i.e., $\tau_c = (X_n^0)^{-1} (x^0)$,
and where we have taken positive ${\dot X}_n^0$. Since this equation is valid at any $x^0$, we omit
in the last step the subscript $|_{\tau = \tau_c}$. Similarly we proceeded in Eq.\,(\ref{3.10}).

Algebraic structure of the constraints (\ref{3.8})--(\ref{3.10}) is independent of foliation of spacetime, determined
by $N$, $N^i$. In any foliation, the form of the equations remain the same. The constraints (\ref{3.8})--(\ref{3.10})
are thus covariant under diffeomorphisms.

By taking a linear combination of those constraints, we obtain a Hamiltonian:
\be
   H = \int \dd^3 \bx \left ( N {\cal H} + N^i  {\cal H}_i \right ) = 0.
\lbl{D1}
\ee
Here ${\cal H} = {\cal H}_G + {\cal H}_m$, ${\cal H}_i={\cal H}_{Gi} + {\cal H}_{mi}$, where ${\cal H}_m$, ${\cal H}_{mi}$,
${\cal H}_G$, and ${\cal H}_{Gi}$ are given in Eqs.\,(\ref{3.9}),(\ref{3.10}), (\ref{3.5})--(\ref{3.6a}).
The terms with $N^i p_{n i}$ cancel out, and so we obtain
\be
  H = \int \dd^3 \bx \, (N\, {\cal H}_{\rm G} + N^i {\cal H}_{{\rm G}i}) + \sum_n p_{n0} = 0.
\lbl{D2}
\ee
In the latter equation, which holds for arbitrary $N$, $N^i$, the matter Hamiltonian is just
the sum of particle's momenta $p_{n0}$, i.e., their energies.

Eq.\,(\ref{D2}) can also be obtained directly from the Einstein equations integrated over a space like hypersurface
according to
\be
  \frac{\kappa}{2} \int \dd \Sigma_\nu \sqrt{-g} \,{G_0}^\nu = - \int \dd \Sigma_\nu \sqrt{-g}\,{T_0 }^\nu 
  = - P_0 .
\lbl{D3}
\ee
If the stress-energy tensor is concentrated so that it effectively forms a discrete
set of particles in the sense as previously described, then the total energy $P_0$ in Eq.\,(\ref{D3})
is just the sum of particle energies, occurring in
Eqs.\,(\ref{D2}).

Instead of Eq.\,(\ref{D3}) we can take the corresponding covariant form
\be
  \frac{\kappa}{2} \int \dd \Sigma_\nu \sqrt{-g}\,{G_\mu}^\nu n^\mu = - \int \dd \Sigma_\nu \sqrt{-g}\,{T_\mu }^\nu n^\mu
  = - P_\mu n^\mu ,
\lbl{D4}
\ee
where $n^\mu$ is a unit vector field. If $T^{\mu \nu} = \rho u^\mu u^\nu$, i.e., a dust stress-energy tensor
and $u^\mu$ the 4-velocity, then it is
convenient to take $n^\mu = u^\mu$. A discrete version of the dust stress energy tensor is the one obtained from
the system of ''point particles", described by the action (\ref{3.2}). Then for $n^\mu$ we can take a vector field that
at the locations of particles coincides with their 4-velocities $u_n^\mu = p_n^\mu/m_n$.

Because the constraints (\ref{3.9})--(\ref{3.10}) are in fact just the the
ADM split of Einstein's equations, we see that as a consequence  of the Bianchi identities they are conserved
in time, or along any time like curve whose tangent is $u^\nu$:
\be
   u^\nu \DD_\nu (G^{\mu \nu} + 8 \pi T^{\mu \nu}) = 0.
\lbl{D5}
\ee
The constraint (\ref{3.8}), namely, $g_{\mu \nu} p_n^\mu p_n^\mu - m_n^2 = 0$, also is conserved,
\be
  u^\alpha \DD_\alpha (g_{\mu \nu} p_n^\mu p_n^\nu - m_n^2) 
  = 2 g_{\mu \nu} u^\alpha p_n^\mu \DD_\alpha p_n^\nu = 2 p_{n \nu} u^\alpha \DD_\alpha p_n^\nu = 0,
\lbl{D6}
\ee
where we now assume that $u^\alpha$ is tangent along a geodesic, so that $p_n^\alpha = m_n u^\alpha$.
Then $u^\alpha \DD_\alpha u^\nu =0$, and the  equality (\ref{D6}) is satisfied.

When quantizing the theory, one has to fix a gauge. In our case this is achieved by  
choice of $N$, $N^i$. Because $N$ and $N^i$ behave as Lagrange multipliers, they can be arbitrary
functions of spacetime coordinates $x^\mu$. 
We will use the choice $N=1$, $N^i =0$. It is straightforward to show
that such gauge fixing imposes no second class constraints\footnote{
See explanation after Eq. (\ref{3.35}).}. 
Upon quantization, in the Schr\"odinger representation in which $X_n^\mu$ and $q_{ij}$ are
diagonal, the momentum operators are the covariant generalizations of
${\hat p}_{n \mu} = - i \p/\p X_n^\mu$ and
${\hat \pi}^{ij} = - i \delta/\delta q_{ij}$. If acting on a functional only once, then ${\hat p}_{n \mu}$ behaves
as partial derivative, and ${\hat \pi}^{ij}$ as functional derivative, otherwise ${\hat p}_{n \mu}$ is
covariant derivative with respect to $g_{\mu \nu}$, and ${\hat \pi}^{ij}$ is covariant functional derivative
with respect to the metric $G_{ijk\ell} \delta (\bx - \bx')$.
The constraints (\ref{3.8})--(\ref{3.10}) become conditions on a state represented as
${\tl \vphi} [T_1,T_2,...T_N,X_1^i,X_2^i,...X_N^i,  q_{ij} (\bx)] {\equiv \tl \vphi} [T_n,X_n^i,q_{ij} (\bx)]$,
$n=1,2,...,\infty$, where $T_n \equiv X_n^\mu$:
\be
  \left ( {\hat p}_{n 0}^2 - q^{ij} {\hat p}_{ni} {\hat p}_{nj}  - m_n^2 \right )
  {\tl \vphi} [T_n,X_n^i,q_{ij} (\bx)] = 0 \hs{1.3cm} , \lbl{3.11} \ee
\be
  \left ( {\cal H}_{\rm G} - \sum_n \delta^3 (\bx - \bX_n) \,i \frac{\p}{\p T_n} \right )
  {\tl \vphi} [T_n,X_n^i,q_{ij} (\bx)] = 0 , \lbl{3.12} \ee
\be
  \left ( {\cal H}_{{\rm G} i} - \sum_n \delta^3 (\bx - \bX_n) \,i \frac{\p}{\p X_n^i} \right )
  {\tl \vphi} [T_n,X_n^i,q_{ij} (\bx)] = 0 . \lbl{3.14} 
\ee

Integrating the last two equations over $\bx$, we obtain
\be
  \left ( H_{\rm G} - \sum_n \,i \frac{\p}{\p T_n} \right ) {\tl \vphi} [T_n,X_n^i,q_{ij} (\bx)] = 0,
\lbl{3.15}
\ee
\be
  \left ( H_{{\rm G} i}- \sum_n \,i \frac{\p}{\p X_n^i} \right ) {\tl \vphi} [T_n,X_n^i,q_{ij} (\bx)] = 0,
\lbl{3.16}
\ee
where\footnote{The ordering issue can be settled by replacing $\delta/\delta q^{ij}$ with covariant functional derivatives with respect to the Wheeler-DeWitt metric, and proceeding \`a la Ref.\,\ci{PavsicOrder}. A further development of this important
point is beyond the scope of the present paper. As in so many other papers,  also here $\delta/\delta q^{ij}$
has only a symbolic meaning, unless considered as the covariant functional derivative. }
\be
  H_{\rm G} = \int \dd^3 \bx \,{\cal H}_{\rm G} =  \int \dd^3 x \, 
    \left ( \frac{1}{\kappa} \, G_{ij\,k \ell} \,
    \frac{\delta^2}{\delta q_{ij} \delta q_{k \ell}}
     + \kappa \, \sqrt{q} R^{(3)} \right ) ,
\lbl{3.17}
\ee
\be
  H_{{\rm G}i} = \int \dd^3 \bx \, {\cal H}_{{\rm G}i} = - \int \dd^3 \bx (-i) 2 \DD_j \frac{\delta}{q_{ij} (\bx)} .
  \hs{2.5cm}
\lbl{3.18}
\ee

At this point let us recall that, as shown by Moncrief and Teitelboim\,\ci{Moncrief}, the momentum
constraint (\ref{3.10}) is a consequence of the conservation of the Hamilton constraint (\ref{3.9}).
Therefore, it is sufficient if we consider the Hamilton constraint (\ref{3.9}) and its quantum versions
(\ref{3.12}) or (\ref{3.15}) only. We then have
\be
  (H_{\rm G} + H_{\rm m}) {\tl \vphi} [T_n,X_n^i,q_{ij} (\bx)] = 0 ,
\lbl{3.19}
\ee
where
\be
   H_{\rm m} = \sum_n {\hat p}_{n 0} = - i \sum_n \frac{\p}{\p T_n} .
\lbl{D5a}
\ee

In this description a quantum state is represented by ${\tl \vphi}[T_n,X_n^i,q_{ij} (\bx)]$, which is
a function of the particles' spacetime coordinates, $X_n^\mu = (T_n,X_n^i)$, and a functional
of the dynamical variables of gravity, $q_{ij} (\bx)$.  In the absence of horizons, which is the
situation that we consider here, we can choose coordinates $X_n^\mu$ so
that all time coordinates $X_n^0 \equiv T_n$
 on a given 3-surface $\Sigma$ are equal: $T_1 = T_2 = ...= T_{\cal N} = T$.
Then we can write ${\tl \vphi} [T_n,X_n^i,q_{ij} (\bx)]$ as a function of a single time coordinate
$T$:
\be
  {\tl \vphi} [T_n,X_n^i,q_{ij} (\bx)] = \phi [T,X_n^i,q_{ij} (\bx)] ~,~~~~~~~n=1,2,...,r,
\lbl{3.22}
\ee
and
\be
  \frac{\dd {\tl \vphi}}{\dd T}= \sum_n \frac{\p {\tl \vphi}}{\p T_n} \frac{\p T_n}{\p T}
 = \sum_n \frac{\p {\tl \vphi}}{\p T_n}  = \frac{\p \phi}{\p T} .
\lbl{3.22a}
\ee
Equation (\ref{3.19}) then becomes
\be
  H_{\rm G} \phi [T,X_n^i,q_{ij}(\bx)]= i \frac{\p \phi [T,X_n^i,q_{ij}(\bx)]}{\p T} .
\lbl{3.23}
\ee

We see that if one starts from the classical action (\ref{3.1a}) in which the matter part is
expressed in terms of the worldlines of individual particles, then upon quantization we arrive
at the Wheeler-DeWitt equation (\ref{3.23}) which contain time. The wave functional
depends on coordinates of particles, $X_n^\mu = (T_n,X_n^i)$, and the induced metric
on a hypersurface $\Sigma$, defined by $X_n^0 =T ={\rm constant}$ (on which all particles have
the same time coordinates $T_n = T$).

Besides the time dependent  equation (\ref{3.23}), the wave functional also satisfies the Klein-Gordon
equation (\ref{3.11}). Assuming {\it real} ${\tl \vphi}$, the second order Klein-Gordon equation
can be cast into the form of a first order equation for a complex wave function\,\ci{Deriglazov,PavsicMiracle}
\be
  {\psi}= \psi_R + i \psi_I ,
\lbl{3.23a}
\ee
where 
\be
  \psi_R = \phi(T,\bX_1,\bX_2,...,\bX_r) = {\tl \vphi} (T_1,T_2,...,T_r,\bX_1,\bX_2,...,\bX_r) ,
\lbl{3.24a}
\ee
and
\be
   \psi_I = i \Omega^{-1} {\dot \phi} (T,\bX_1,\bX_2,...,\bX_r) .
\lbl{3.24b}
\ee
So instead of (\ref{3.11}) we obtain the equivalent equation (see Appendix B)
\be
 i \frac{\p {\psi} }{\p T} = \Omega \psi .
\lbl{3.25}
\ee
Here $\Omega$ is a matrix, by means of which the above equation is just a compact form
of Eq.\,(\ref{B4a}) that we derived in Sec.\,2
after expanding the Schr\"odinger wave functional $\Phi [t,\vphi (\bx)]$ in terms
of multiparticle wave functions according to Eq.\,(\ref{4.19}), and after generalizing
it so to include the 3-surface induced metric $q_{ij}$ as well.

The function $\psi$, occurring in Eq.\,(\ref{3.25}), is the true, complex valued, wave function (related to the
probability density), whilst the function ${\tl \vphi}$ satisfying the Klein-Gordon equation
(\ref{3.11}), is just a real field.

By using the expression (\ref{3.23a}) and Eq.\,(\ref{3.23}}), it is straightforward to derive that in addition
to Eq.\,(\ref{3.23}), valid for the real field ${\tl \vphi}$, we also have the similar equation for the
complex wave function $\psi$:
\be
   H_{\rm G} \psi [T,X_n^i,q_{ij}(\bx)]= i \frac{\p \psi [T,X_n^i,q_{ij}(\bx)]}{\p T} .
\lbl{3.26}
\ee

In the above treatment we  started from the classical action (\ref{3.1a})
and the corresponding constraints which upon quantization became the conditions
on states (\ref{3.11})--(\ref{3.14}) that can be represented either as a wave functional
${\tl \vphi}[T_n,X_n^i,q_{ij} (\bx)]$ or ${\phi}[T,X_n^i,q_{ij} (\bx)]$.
No further, i.e., a ``second'' quantization is performed here. The complex
functionals  $\psi[T,X_n^i,q_{ij} (\bx)]$, $n=1,2,3,...$, are obtained according
to the prescriptions (\ref{3.23a})--(\ref{3.24b}).
The matter  Hamiltonian $H_{\rm m}$
plus the gravitational Hamiltonian $H_{\rm G}$ acting together on $\psi[T,X_n^i,q_{ij} (\bx)]$
 give zero. Time automatically appears in the equations, such as (\ref{3.25}) or (\ref{3.26}).

\subsection{Multiparticle states arising from the traditional approach, and their relation
to the novel approach}

In the previous subsections we discussed a novel approach in which we started from a classical
matter represented as a multi particle system.
In the usual treatments matter action is not $I_m [X_n^\mu, g_{\mu \nu}(x)]$, or
equivalently, $I_{\rm m} [X_n^\mu,p_{n \mu},\lambda_n , g_{\mu \nu}]$
(Eqs.\,(\ref{3.2}),(\ref{3.3})), but is a functional of fields, for instance scalar fields:
\be
  I_{\rm m} [\vphi^a (x),g_{\mu \nu} (x)] = \frac{1}{2} \int \dd^4 x\,\sqrt{-g}\,
  \left ( g^{\mu \nu} \p_\mu \vphi^a \p_\nu \vphi_a - m^2 \vphi^a \vphi_a \right ) ,
\lbl{3.27}
\ee
which after the ADM split reads:
\bear
  &&I_{\rm m} [\vphi^a,q_{ij},N,N^i] = \frac{1}{2} \int \dd t \, \dd^3 \bx N \sqrt{q}
  \left [ \frac{1}{N^2} ({\dot \vphi}^a - N^i \p_i \vphi^a )  ({\dot \vphi}_a - N^j \p_j \vphi_a ) \right .
  \nonumber \\
 &&\hs{7cm}  - q^{ij} \p_i \vphi^a \p_j \vphi_a - m^2 \vphi^a \vphi_a \biggr] .
\lbl{3.28}
\ear

Variation of  the total action
\be
  I = I_{\rm m} [\vphi^a,q_{ij},N,N^i] + I_{\rm G} [q_{ij},N,N^i]
\lbl{3.30}
\ee
with respect to $N$ and $N^i$ gives the constraints ${\cal H} = {\cal H}_{\rm G} + {\cal H}_{\rm m}$
and ${\cal H}_i = {\cal H}_{{\rm G}i} + {\cal H}_{{\rm m}i}$. Here the gravitational part of
the constraints, ${\cal H}_{\rm G}$, ${\cal H}_{{\rm G}i}$, are given in Eqs.\,(\ref{3.5}),(\ref{3.6}),
whilst the matter parts ${\cal H}_{\rm m}$, ${\cal H}_{{\rm m}i}$ are now
\bear
  &&{\cal H}_{\rm m} = \frac{\sqrt{q}}{2} \left ( \frac{\Pi^a \Pi_a}{q} + q^{ij} \p_i \vphi^a \p_j \vphi_a
  +m^2 \vphi^a \vphi_a \right ) \lbl{3.31}\\
  &&{\cal H}_{{\rm m}i} = \p_i \vphi^a \Pi_a ,
\lbl{3.32}
\ear
where
\be
  \Pi_a = \frac{\p {\cal L}_{\rm}}{\p {\dot \vphi}^a}
   = \frac{\sqrt{q}}{N} ({\dot \vphi}_a - N^i \p_i \vphi_a) .
\lbl{3.33}
\ee

The Hamiltonian is a linear combination  of the constraints
\be
  H = \int \dd^3 \bx (N {\cal H} + N^i {\cal H}_i ) .
\lbl{3.34}
\ee
The Lagrange multipliers $N$ and $N^i$ are arbitrary, and, as before, we choose $N=1$,
$N^i=0$, so that
\be
  H = \int \dd^3 \bx \, {\cal H} = \int \dd^3 \bx \, ({\cal H}_{\rm G} + {\cal H}_{\rm m})
  = H_{\rm G} + H_{\rm m} .
\lbl{3.35}
\ee
Such a choice brings no second class constraints, because it involves only Lagrange multipliers and their
conjugate momenta $\pi$, $\pi_i$, and no other phase space variables. Namely, setting
$\phi_1 = N-1$, $\phi_2 = \pi$, $\phi_1^i = N^i$, $\phi_{2i}= \pi_i$, and calculating the time derivative
of the constraints according to ${\dot \phi}= \lbrace \phi,H \rbrace$, we obtain
${\dot \phi}_1 = 0$,  ${\dot \phi}_2 = - {\cal H}$, ${\dot \phi}_1^i = 0$, ${\dot \phi}_{2 i} = {\cal H}_i$,
which are the original constraints.

Upon quantization we have
\be
  (H_{\rm G} + H_{\rm m})|\Phi \rangle = 0 ,
\lbl{3.36}
\ee
which in the Schr\"odinger functional representation gives
\be
  \langle \vphi^a (\bx),q_{ij} (\bx)|(H_{\rm G} + H_{\rm m})|\Phi \rangle = 0.
\lbl{3.37}
\ee

Explicitly, the latter equations reads
$$ \int \dd^3 \bx \left [
  \frac{1}{\kappa} G_{ijk \ell} \frac{\delta^2}{\delta q_{ij} \delta q_{k \ell}} + \sqrt{q} R^{(3)}
  \right . \hs{4.5cm} $$
\be
  \hs{2cm}+ \left . \frac{\sqrt{q}}{2} \left (-\frac{\delta^2}{\delta \vphi^a \delta \vphi_a}+ \vphi^a( -\DD^i \DD_i  
  +m^2) \vphi_a  \right ) \right ] \Phi [\vphi^a (\bx),q_{ij} (\bx)] = 0 ,
\lbl{3.38}
\ee
where  ${H}_{\rm G}$,  ${H}_{\rm m}$ are represented as matrices in the
space of fields $\vphi^a (\bx)$, $q_{ij} (\bx)$, i.e., as functional differential operators.
Concerning the factor ordering of the operators ${\hat \pi}^{ij}$ we can adopt and generalize
the procedure of Ref.\,\ci{PavsicOrder}.

The functional $\Phi [\vphi^a (\bx),q_{ij} (\bx)]$
is ``first quantized'' with respect to the metric $q_{ij} (\bx)$, and ``second quantized'' with respect to
particle position. This is a ``hybrid'' procedure, and there is manifestly no time in Eq.\,(\ref{3.38}). 

The matter Hamiltonian in Eq.\,(\ref{3.38}),
\be
   H_m = \int  \dd^3 \bx \, \frac{{\sqrt q}}{2} \left ( - \frac{\delta^2}{\delta \vphi^a \vphi_a}
   + \vphi^a \left ( - \DD_i \DD^i + m^2 \right ) \vphi_a \right ),
\lbl{D9}
\ee
 can be expressed in terms of the operators
$a[\bx,q_{ij} (\bx)]$, $a^\dg[\bx,q_{ij} (\bx)]$, defined in Sec.\,2, so we have
\be
  {H}_{\rm m} = \int \dd^3 \bx \left ( a^\dg[\bx,q_{ij} (\bx)] \sqrt{m^2 - \DD^i \DD_i} 
 \,  a [\bx,q_{ij} (\bx)] + {\rm z. p.} \right ).
\lbl{3.39}
\ee
This is an exact expression for ${H}_{\rm m}$.

Let us now take the following Ansatz:
\be
  \Phi[\vphi^a (\bx),q_{ij} (\bx)] = \sum_n \dd^3 \bx_1 ... \dd^3 \bx_n\,
  \psi [\bx_1,...,\bx_n,q_{ij} (\bx)] {\tl \Phi}_n [\vphi (\bx),q_{ij} (\bx)]  ,
\lbl{3.40}
\ee
where 
\be
{\tl \Phi}_n [\vphi (\bx),q_{ij} (\bx)] = a^\dg [\bx_1,q_{ij}(\bx) ...a^\dg [\bx_n ,q_{ij}(\bx) {\tl \Phi}_0 [\vphi,q_{ij} (\bx)]
\lbl{3.40a}
\ee
is obtained by the action of the creation operators on the vacuum ${\tl \Phi}_0[\vphi,q_{ij} (\bx)]$,
which is now not the vacuum (\ref{4.q4}), valid in the case of a fixed background metric, but a suitably generalized expression accounting for the fact that now the metric is dynamical. Formally, let us set
\be
{\tl \Phi}_0[\vphi,q_{ij} (\bx)] \propto {\rm exp} 
\left [-\frac{1}{2}  \int \dd^3 \bx {\sqrt q} \vphi (\bx) \sqrt{m^2-D^i D_i} \,\vphi +
Q[q_{ij}] \right ],
\lbl{3.40b}
\ee
 which contains in the exponential an additional functional $Q[q_{ij}]$, that could contain, among others,
the expressions such as $\int \dd^3 \bx \sqrt{q} R^{(3)}$.

Using the commutation relations\footnote{
As we pointed in Sec.\,2, the metric and its determinant cancel out, so that they do not appear in the
r.h.s. of the commutation relation.}
\be
  [a[\bx,q_{ij} (\bx)], a^\dg [\bx',q_{ij} (\bx)]] = \delta^3 (\bx - \bx'),
\lbl{3.41a}
\ee
\be
  [a[\bx,q_{ij} (\bx)], a [\bx',q_{ij} (\bx)]] = 0~,~~~~ [a^\dg [\bx,q_{ij} (\bx)], a^\dg [\bx',q_{ij} (\bx)]] = 0,
\lbl{3.41b}
\ee
the expansion (\ref{3.40}), and the matter Hamiltonian (\ref{D9}), we find
\bear
  &&H_{\rm m} \Phi = \sum_{r=1}^\infty \int \dd^3 \bx_1 \dd^3 \bx_2 ... \dd^3 \bx_n \sum_{n=1}^r
  \sqrt{m_n^2 - \DD_n^2 }\, \psi [\bx_1,\bx_2,...,\bx_r , q_{ij} (\bx) ] \nonumber\\
  && \hs{2cm} \times \, a^\dg [\bx_1 , q_{ij} (\bx)] ...
  a^\dg [\bx_r, q_{ij} (\bx)] \Phi_0 [\vphi(\bx),q_{ij} (\bx)] ,
\lbl{3.42}
\ear
where
\be
  \DD_n^2 \equiv  q^{ij} \DD_{ni} \DD_{nj}~ , ~~~~~~\DD_{ni} \equiv \frac{\DD}{\DD X_{ni}} .
\lbl{3.42a}
\ee

If we postulate that the quantizations based 
on the actions (\ref{3.1a}) and (\ref{3.27}) are equivalent, then the wave functional
$\psi[\bx_1,...,\bx_n,q_{ij} (\bx)]$ is the same one as considered
in Eqs.\,(\ref{3.11})--(\ref{3.16}), depends on time $T$ and satisfies (\ref{3.26}).  Here we have
renamed $X_n^i \equiv {\bX}_n$, occurring in Eqs.\,(\ref{3.11})--(\ref{3.16}), into $\bx_n$, $n=1,2,...,{\cal N}=\infty$.
If  time $T$ occurs in $\psi$, then it automatically also occurs in $\Phi$.

Taking now into account Eq.\,(\ref{3.26}), we obtain
\be
 H_{\rm m} \Phi  =i \frac{\p \Phi}{\p T}  ,
\lbl{3.46}
\ee
We have thus reproduced  the functional representation of the time dependent Schr\"odinger equation for the scalar field in 
the presence of a 3-metric field $q_{ij}(\bx)$.

Returning to Eq.\,(\ref{3.37}) and using the latter result, namely that $H_{\rm m} \Phi = i \p\Phi/\p T$,
we obtain
\be
  H_{\rm G} \Phi = - i \frac{\p \Phi}{\p T} .
\lbl{3.48}
\ee
We have thus obtained a time dependent Schr\"odinger equation for the gravitational part
of the Hamilton operator.  Both equations, (\ref{3.46}) and (\ref{3.48}), together give the Wheeler-DeWitt equation
(\ref{3.37}). It thus turns out that though time does not manifestly take place in the Wheeler-DeWitt equation
(\ref{3.37}), it is hidden in the wave functional $\Phi$, if  the quantization procedures discussed
in Secs.\,(3.1)and (3.2) are equivalent in the sense that they lead to the same physics.

In the approach considered in this paper the problem of time does not exist, because we started from
the classical action in which gravity is coupled to a multi-particle system that by definition contains time;
upon quantization of such system, time does not disappear. This is different from the usual approaches
in which the  classical action to be quantized contains gravity coupled to a field, for instance a scalar field,
and there is a problem of how to assign the role of time to certain degrees of freedom entering
the quantum equations.

There are many different approaches to the problem of time in quantum gravity, extensively reviewed
by  Anderson\ci{Anderson}. In particular, Lapchinsky and Rubakov\ci{Lapchinsky}  showed that it is possible to introduce
a Tomonaga bubble-time parameter $\sigma (x)$ denoting spatial hypersurfaces that spacetime is foliated to.
In their approach, the gravity part of the Hamilton constraint gives in the case where $\sigma = const = t$,
 the time derivative of the matter state in the presence of a quasiclassical background gravitational field.

Such an approach that requires consideration of the quasiclassical approximation was criticized
by Barvinsky\ci{Barvinsky}, because it was not just a conventional calculational tool, but the only
means to introduce the notion of time, probability, etc., in quantum cosmology. Therefore, Barvinsky
developed a procedure that employed no (quasiclassical) approximation, but introduced spacetime
foliation by imposing gauge conditions on the degrees of freedom of gravity and matter. In that approach,
matter degrees of freedom were represented by scalar fields. But in principle an analogous procedure
holds for the case in which matter degrees of freedom are represented by multi-particle spacetime
coordinates $X_n^\mu$ occurring in the action\,(\ref{3.3}). A peculiar feature of such approach is
that the time-like coordinates $X_n^0 \equiv T_n$ represent just the many-fingered time associated
with a spacetime foliation. A particular choice of coordinates, such that $T_1 =T_2=...= T_n =T$
(see Eq.\,(\ref{3.22})) corresponds to $\sigma = const =t$, mentioned above. The many-fingured time
thus occurs in the very construction of the matter action. Therefore, the gauge condition that
reduces the number of variables to the physical degrees of freedom can be $N=1$, $N^i =0$
and is straightforwardly associated with a particular spacetime foliation in the quasiclassical limit.

An alternative procedure was considered by Peres\ci{Peres} who transformed a constrained system to
an unconstrained one by a suitable canonical transformation of the phase space variables, such that one of
the so obtained variables serves the role of time. No such a canonical transformation is necessary
in our approach in which the coordinates $X_n^0 \equiv T_n$ already have the role of (many-fingered) time.

%%%%%%%%%%%%%%%%%%%%%%%%%%%%%%%%%%%%%%%%%%%%%%%%

\section{Discussion}

When investigating the universe and behaviour of objects in it, we normally
have their positions in mind. In quantum gravity the entire universe is
envisaged to be describable by a wave function.
However, in practical calculations only few degrees of freedom of the
universe are taken into account, but we assume that in reality there exists a wave function(al)
of the universe that comprises all degrees of freedom, including those of observers.
We base our discussion of the universe on canonical gravity which we modified in the part
that relates to matter. Usually matter is represented by fields, such as a scalar, spinor, gauge field, etc.
In the model considered in this paper we started directly from the classical action for a multi particle
system coupled to gravity, and arrived after quantization to the Schr\"odinger equation for
a set of multi-particle wave functionals $\psi[t,\bx_1,\bx_2,...,\bx_r, q_{ij} (\bx)]$, $r=1,2,...\, .$.  
Thus, in our approach a state satisfying the Wheeler-DeWitt equation is represented
as a functional of gravity and matter degrees of freedom, the latter being
given in terms of multi particle configurations. It then turns out that time does not
disappear from the quantum equations, it is given in terms  of the time like coordinates
of particles, associated with the clocks situated on particle's worldlines.
 We have thus set up a theoretical framework for a
quantum gravity description of the universe, which involves positions of particles, a concept
in many respect closer to our intuition and observations than the concept of scalar
field. 

In this paper we have confined us to discussing the Wheeler-DeWitt equation whose semiclassical
solutions are well-known to contain singularities. How to avoid singularities has been discussed,
e.g., by Claus Kiefer and Barbara Sandhofer\ci{KieferSingularity} (see also refs\,\ci{KieferSingularity[6],KieferSingularity[2],KieferSingularity[34],KieferSingularity[43],KieferSingularity[28]}
cited therein) who conclude:
 "Upon discussing the Wheeler-DeWitt equation, one finds that all normalizable solutions
lead to a wave function that vanishes at the point of the classical singularity;
this we interpret as singularity avoidance. An analogous situation of singularity avoidance
is found in the loop quantum cosmology of this model\,\ci{KieferSingularity[39]}." Our classical action (\ref{3.1a})
 of point particles coupled to gravity, if taken literally, would be impossible,
 because at the positions $X_n^\mu$
of point particles there would be black hole singularities. We avoided singularity by postulating that
the particles are extended and that $X_n^\mu$ are coordinates of their effective positions (analogous
to the center of mass of a non relativistic particle). 

Our procedure could as well be upgraded into a promising direction: namely, by extending
the classical point particle to the superparticle, by including, besides the commuting
coordinates $X^\mu$, also the Grassmann anticommuting coordinates. and coupling such
system to gravity. Thus, because of the presence of the additional, anticommuting coordinates,
$\xi^\mu$, the problem of a point particle coupled to a gravitational field would be avoided,
because the extra coordinates would bring into the game in an elegant way the particle's
effective extension. Upon quantization of such a model in which a superparticle is coupled
to the metric (which now depends on $X^\mu$ and $\xi^\mu$), spinor fields would occur
in the description.  For the time being, in our current paper we describe a classical particle by
four coordinates $X^\mu$ only, and assume that it is extended and held from collapsing into a black
hole by the forces that are not included into the description. In our opinion this is legitimate,
because a physical {\it model} necessarily involves a restricted set of variables, fields, etc.,
and neglects the rest.

However, we have to bear in mind that the action of Einstein's gravity containing only the first order
term of the curvature scalar, $R$,  leads upon quantization to divergences that cannot be renormalized.
In the present paper we have not addressed this problem. One possibility is to  include into
the gravity part of the action (\ref{3.1a}) the quadratic, $R^2$, and higher order/derivative terms as well, and
re-run the procedure considered in this paper, either with the same matter action consisting of point
particles, or generalizing them to superparticles, as mentioned above. Inclusion of higher
derivative terms would then solve the problem of renormalizability. But, as widely
recognized, we would then have problems with ghosts and instabilities. There is a vast literature on how
higher derivative gravity theories can be made physically viable (see, e.g., \ci{PavsicPUReview}, and references
therein.)

A higher derivative gravity action arises as an effective action of string theory..
Another promising direction of research is loop quantum gravity.
Both those theories have
satisfactorily addressed the issue of quantum divergences. But string theory has run
into serious problems, including the so called ``landscape", while in loop quantum gravity it is
not yet clear how spacetime manifold emerges  from such a setup.
I anticipate that the different theoretical structures, such as a higher derivative canonical
gravity with superparticles as sources (that can be extended to superstrings), string theory, and
loop quantum gravity,  will at the end turn out to be revelling different
aspects of an underlying more fundamental theory that is being crystalized in numerous works based
on the powers of Clifford algebras.

\vs{8mm}

\centerline{\large \bf Acknowledgement}

\vs{2mm}

This work has been partly supported by the Slovenian Research Agency.
 \vs{1.5cm}

{\large \bf Appenix A: Coupling of ``point-like'' sources to the gravitational field}

\vs{2mm}

Because classical gravity contains black hole solutions, point particle sources are
problematic\footnote{See the paper\,\cite{Geroch}, where those problems are
thoroughly analysed not only for point particles, but also for branes.}. However,
if a particle is extended beyond its Schwarzschild radius, in principle there is no
problem. When describing motion of an extended object, we may neglect its
internal dynamics and consider only the motion of an effective worldline, a ``center of mass''.
In such a
case the coupling of the object with the gravitational field may be given by Eq.\,(\ref{3.3}),
with understanding that $X^\mu (\tau)$ are center of mass coordinates, and that the
range of the considered spacetime coordinates is outside the Schwarzschild radius.

As a model of extended object let us consider an open\footnote
{This means that the
brane does not fill the entire space, but only the volume inside a sphere, so that
our brane is in fact a ball-filling brane.}
space filling brane, described by coordinates
$X^\mu (\xi^A)$, $\mu = 0,1,2,...,D-1$, $A=0,1,2,...p$, where in our case of
a space filling brane, $p=D-1$. The action for such a system is
\be
I_{\rm m} = \mu_B \int \dd^{p+1} \xi \, (-{\rm det} \p_A X^\mu \p_B X_\mu )^{1/2}~,
~~~~~~~f_{AB} \equiv \p_A X^\mu \p_B X_\mu ,
\lbl{AA1}
\ee
where $\mu_B$ is the brane tension. When considered as a matter source of
gravity the above action has to be rewritte so to include a $\delta$-function:
\be
I_{\rm m} = \mu_B \int \dd^{p+1} \xi \, (-{\rm det} (\p_A X^\mu \p_B X_\mu )^{1/2}
    \delta^D (x-X(\xi)) \dd^D x .
\lbl{AA2}
\ee
This action is invariant under reparametrizations of the worldsheet parameters
$\xi^A$, and under general coordinate transformations of spacetime coordinates
$x^\mu$. Rewritten in terms of new $\xi^A$ and $x^\mu$, it retains the same
form (\ref{AA2}), but with new functions $X^\mu (\xi^A)$, representing the same
worldsheet.

Let us now split the derivative according to\ci{Barut,PavsicPointLike}
\be
  \p_A = n_A \p + {\bar \p}_A,
\lbl{AA10}
\ee
where $n_A$ is a vector in the direction of the hypersurface element $\dd \Sigma_A = \dd \Sigma n ^A$,
 and $\p$ is the derivative in the direction of $n_A$
(normal derivative), whilst ${\bar \p}_A$ is the tangential derivative,
orthogonal to $n_A \p$. Then we have
\be
   f_{AB} = \p_A X^\mu \p_B X_\mu = n_A n_B \p X^\mu \p X_\mu +
   {\bar \p}_A X^\mu {\bar \p}_B X_\mu ,
\lbl{AA12}
\ee
from which we obtain
\be
   n^A n_A = \frac{1}{\p X^\mu \p X_\mu} ,
\lbl{AA13}
\ee
Inserting (\ref{AA12})
into the definition of $f = {\rm det} f_{AB}$, we obtain\,\ci{Barut,PavsicPointLike}
\be
   f = \frac{{\tl f}}{n^2}~,~~~~~~{\tl f} = \frac{1}{p!} \epsilon^{A_1 ...A_p}
    \epsilon^{B_1 ...B_p} {\bar f}_{A_1 B_1} ... {\bar f}_{A_p B_p}  ~,
    ~~~~~~{\bar f}_{A B} \equiv {\bar \p}_A X^\mu {\bar \p}_B X_\mu .
\lbl{AA14}
\ee

In a gauge in which $n_A = (1,0,0,...)$, using the procedure of Ref.\,\ci{PavsicPointLike}, we have
$\xi^A = (\tau,\sigma^a)$, $a=1,2,...,p$, $\p X^\mu = \p/\p \tau \equiv {\dot X}^\mu$,
${\tl f} = {\rm det} f_{ab} \equiv {\bar f}$, $\p_a = \p/\p \sigma^a$, and $\dd \Sigma = \dd^p \sg$.

\begin{comment}
If the brane is a ball filling brane, then we can choose a gauge
such that the velocity ${\dot X}^\mu (\tau,\sg)$ is independent of $\sg$, so that
$X^\mu (\tau,\sg) = X_{\rm T}^\mu (\tau) + w^\mu (\sg)$. Then the 4-velocity is the
same across the ball, ${\dot X}^\mu (\tau,\sg) = {\dot X}_{\rm T}^\mu (\tau)$,
where ${ X}_{\rm T}^\mu (\tau)$ is the center of mass worldline.
\end{comment}

Using (\ref{AA10})--(\ref{AA14}), the action (\ref{AA2}) then reads
\be
  I_{\rm m} = \mu_B \int \dd \tau\, \dd^p \sg \, \sqrt{-{\tl f}} \sqrt{ \p X^\mu \p X_\mu}
  \,\delta^D (x- X(\tau,\sg)) \dd^D x, 
\lbl{AA15}
\ee
which is a covariant expression, because it does not change
its form under the transformations $\xi^A \rightarrow \xi'^A = \chi^A (\xi)$ and
$x^\mu \rightarrow x'^\mu =F^\mu (x)$.
In a particular gauge (choice of parameters $\xi^A$), considered above,
the action becomes
\be
  I_{\rm m} = \mu_B \int \dd \tau\, \dd^p \sg \, \sqrt{-{\bar f}} \sqrt{ {\dot X}^\mu {\dot X}_\mu}
  \,\delta^D (x- X(\tau,\sg)) \dd^D x, 
\lbl{AA16}
\ee

Let us now choose a line $X_{\rm T}^\mu (\tau)$ and write
\be
  X^\mu (\tau,\sg) = X_{\rm T}^\mu (\tau) + w^\mu (\tau, \sg).
\lbl{AA17}
\ee
Then
\be
  I_{\rm m} = \mu_B \int \dd \tau\, \dd^p \sg \, \sqrt{-{\bar f}} \sqrt{ {\dot X}_{\rm T}^\mu {\dot X}_{{\rm T}\mu}}
  \,\delta^D (x- X_{\rm T} (\tau) - w^(\tau,\sg)) \dd^D x ,
\lbl{AA18}
\ee
where we have now ${\bar f} = {\rm det}\, \p_a w^\mu \p_b w_\mu$.

Next, let us use the expansion\footnote{
We can verify on a simpler example that such expansion indeed works:
$$\int  \dd x \, F(x) \delta (x-a)  = F(-a)$$
$$\int \dd x\, F(x) \left (\delta (x) + \frac{\p \delta (x-a)}{\p a}\bigg|_{a=0} a + ... \right )
= F(0) - F' (0) a + ... = F(-a)$$.},
$$
  \delta (x-X_{\rm T} - w(\tau,\sg)) = \delta (x-X_{\rm T} (\tau) )  \hs{6cm}$$
\be
 \hs{1cm}+\int \dd^p \sg w^\mu(\tau,\sg)\frac{\delta \, \delta \left ( x-X_{\rm T} (\tau) 
 - w(\tau,\sg) \right )}{\delta w^\mu (\tau,\sg)}\bigg|_{w(\tau,\sg)=0}+...,
\lbl{AA19}
\ee
 and insert it into (\ref{AA18}). Taking also a gauge such that the determinant ${\bar f}$ does not depend on $\tau$,
the quantity $m= \mu_B \int \dd^p \sg \sqrt{-{\bar f}}$ can be factored out. Assuming that the size of the brane is small and
neglecting the terms with the powers of $w^\mu$, we obtain
\be
  I_{\rm m} = m \int \dd \tau \sqrt{{\dot X}_{\rm T}^\mu {\dot X}_{{\rm T} \mu}}\,
   \dd^D x \sqrt{- g}  \,\frac{\delta^D (x - X_{\rm T} (\tau)}{\sqrt{-g}}
 ~,~~~~~~x \notin \Omega .
\lbl{AA20}
\ee
In the latter equation we have effectively approximated
$X^\mu (\tau, \sg)$ with $X_{\rm T} (\tau)$, and taken only the region
$x \notin \Omega$ of the spacetime outside the ball.

The action (\ref{AA20}) implies that $X_{\rm T} (\tau)$ is a geodesic. This is consistent with the brane equations of
motions corresponding to the action (\ref{AA1}), $  \DD_A \DD^A X^\mu =0$, which can be split as
$\DD_\tau \DD^\tau X_{\rm T} + \DD_\tau \DD^\tau w + \DD_a \DD^a w^\mu = 0$. The $w$-terms in the latter equation are
due to the extension of the object and represent a deviation from the geodesic equation, like in the Papapetrou
equation. If we neglect them, then we have the geodesic equation.

The above example of a ball, modelled \`a la space filling brane, shows how an extended object can be
approximately described as a point particle coupled to the gravitational field, with understanding that only the
region outside the horizon is taken into account. The region inside horizon is not taken into account, because
the object is actually not point-like, but extended. With the ball, we do not have a smeared $\delta$-function,
but a ``true'' $\delta$ function, namely, $\delta^D (x - X(\tau, \sg))/\sqrt{-g}$, i.e., an object which is covariant
under general coordinate transformations of $x^\mu$, and also under reparametrization of $\xi^A = (\tau,\sg^a)$.

The procedure explained above can be straightforwardly adapted to hold not only for a brane, filling a ball, but also for
any brane, for instance, for a closed 2-brane, considered by Dirac as a model of electron.

In our procedure we in fact avoided the problem of defining the center of mass in special an general relativity.
Namely, a far away observer cannot distinguish among arbitrarily chosen lines within the extended object whose size
is negligible in comparison with the considered distances. Let us now nevertheless demonstrate how the center
of mass could be defined, first in flat spacetime and then in a curved one.

Choosing a unit time like direction $n^\mu$ in Minkowski space, let us define the center of mass coordinates
for a system of point particles according to
\be
  X_{\rm T}^\mu = \frac{\sum_k p_k^\rho n_\rho X_k^\alpha {N_\alpha}^\mu}{\sum_k p_k^\sg n^\sg} ,
\lbl{AA25}
\ee
where ${N_\alpha}^\mu = {\delta_\alpha}^\mu - n_\alpha n^\mu$ is the projector onto the hypersurface $\Sigma_\mu$, orthogonal to
$n_\mu$. The center of mass coordinates can thus be interpreted as being defined with respect to a
chosen simultaneity surface $\Sigma_\mu$, associated with an observer.

The Poisson brackets between so defined center of mass coordinates and the total momentum
$P^\nu = \sum_k P_k^\nu$ are
\bear
  &&\lbrace X_{\rm T}^\mu,X_{\rm T}^\nu \rbrace = \frac{\p X_{\rm T}^\mu}{\p X_k^\beta}\frac{\p X_{\rm T}^\nu}{\p p_{k \beta}}
  -  \frac{\p X_{\rm T}^\nu}{\p X_k^\beta}\frac{\p X_{\rm T}^\mu}{\p p_{k \beta}} = 0 , 
\lbl{AA26a}\\
 && \lbrace X_{\rm T}^\mu,P^\nu \rbrace = N^{\mu \nu} = \eta^{\mu \nu} - n^\mu n^\nu .  
\lbl{AA26b}
\ear
These equations are Lorentz covariant. In the particular case of $n^\mu = (1,0,0,0)$, we have
\bear
  &&\lbrace X_{\rm T}^0,P^0 \rbrace = \eta^{0 0} - 1 = 0 ,  \lbl{AA27}\\
  &&\lbrace X_{\rm T}^r,P^s \rbrace = \eta^{rs}~ ,~~~~~~r,s = 1,2,3 \lbl{AA28}\\
  &&\lbrace X_{\rm T}^0,P^s \rbrace = \eta^{0 s} - n^0n^s = 0, \lbl{AA29}\\
  &&  X_{\rm T}^\mu = \frac{\sum_k p_k^0 (X_k^\mu - X_k^0 n^\mu)}{\sum_k p_k^0}=
    \left\{
    \begin{array}{rl}
      0 ~~~~& \text{if } \mu=0,\\
      \frac{\sum_k p_k^0 X_k^r}{\sum_k p_k^0}& \text{if } \mu =r.
    \end{array} \right. \lbl{AA30}
\ear
We see that these are correct Poisson bracket relations.

For a generic stress-energy tensor we have
\be
  X_{\rm T}^\mu = 
    \frac{\int \dd \Sigma_\nu T^{\nu \rho} n_\rho X^\alpha {N_\alpha}^\mu}{\int \dd \Sigma_\nu T^{\nu \rho} n_\rho}.
\lbl{AA31}
\ee

How to define the center of mass in curved spacetime is much debated, with no unique generally accepted
solution. Our tentative proposal is first to consider $x^\alpha$ as a vector field which in given coordinates\,\ci{PavsicBook}
is $a^\alpha (x) = x^\alpha$. Then we can generalize (\ref{AA31}) to
\be
  X_{\rm T}^\mu = 
    \frac{\int^{[x_0]} \dd \Sigma_\nu \sqrt{-g} \,T^{\nu \rho} n_\rho a(x)^\alpha {N_\alpha}^\mu}
    {\int \dd \Sigma_\nu \sqrt{-g}\,T^{\nu \rho} n_\rho}.
\lbl{AA32}
\ee
where $\int^{[x_0]}$ denotes the covariant integral over a vector field, which in the above case is
$A^\mu (x) = a^\alpha {N_\alpha}^\mu$. This means that the vectors $A^\mu (x)$ at different
points $x$ are paralelly transported along a geodesic from the point $x$ to a chosen point $x_0$
(the ``origin''), where they are summed (integrated). How precisely this works is shown in
Refs.\,\ci{Folomeshkin,LogunovFolomeshkin} and \ci{PavsicOrder}.

For the purpose of  the procedure adopted in this paper it is sufficient that the center of mass, or any
other point that samples the motion of a finite size particle, does exist. The precise location of such point
within the particle is not important for the validity of our procedure.

\newpage

{\large \bf Appenix B: The connection between the multiparticle Schr\"odinger and Klein-Gordon equation}

\vs{2mm}

The multiparticle Schr\"odinger equation (\ref{B4a}) can be cast into the following system of
equation for the real and the imaginary part of the complex wave function $\psi = \psi_R + i \psi_I$:
\be
  \frac{\p \psi_R}{\p t}= \sum_{k=1}^r \om_{\bx_k} \psi_I (t,\bx_1,\bx_2,...,\bx_r) ,
\lbl{B1}
\ee
\be
  \frac{\p \psi_I}{\p t}= -  \sum_{k=1}^r \om_{\bx_k} \psi_R (t,\bx_1,\bx_2,...,\bx_r) ,
\lbl{B2}
\ee

Introducing the compact notation
\be
   \bx_1,\bx_2,...,\bx_r \equiv \bX_r~,~~~~~~~\psi_R (\bx_1,\bx_2,...,\bx_r)\equiv  \psi_R(\bX_r) \equiv \psi_R^{(\bX_r)},
\lbl{B3}
\ee
\be
  {\Omega^{(\bX_r)}}_{(\bX_s)} = \sum_{k=1}^r \om_{\bx_k} {\delta^r}_s \delta (\bX_r - \bX'_s) ~,
  ~~~~\om_{\bx_k} = \sqrt{m^2 - \DD_{\bx_k}} .
\lbl{B4}
\ee
we can rewrite Eqs.\,(\ref{B1}),(\ref{B2}) as
\be
  {\dot \psi}_R^{(\bX_r)} = {\Omega^{(\bX_r)}}_{(\bX_s)} \psi_I^{(\bX_s)} ,
\lbl{B5}
\ee
\be
  {\dot \psi}_I^{(\bX_r)} = -  {\Omega^{(\bX_r)}}_{(\bX_s)} \psi_R^{(\bX_s)} ,
\lbl{B6}
\ee
Expressing $\psi_I^{(\bX_s)}$ in Eq.\,(\ref{B5}) in terms of ${\dot \psi}_R^{(\bX_r)}$,
\be
  \psi_I^{(\bX_s)}= {{\Omega^{-1}}^{(\bX_r)}}_{(\bX_s)} {\dot \psi}_R^{(\bX_r)},
\lbl{B7}
\ee
and inserting it into Eq.\,(\ref{B6}), we obtain the following second order equation:
\be
  {\ddot \psi}_R^{(\bX_r)} 
  + {\Omega^{(\bX_r)}}_{(\bX_s)} {\Omega^{(\bX_s)}}_{(\bX_k)} \psi_R^{(\bX_r)} .
\lbl{B8}
\ee
Taking into account the explicit form (\ref{B4}) of the matrix ${\Omega^{(\bX_r)}}_{(\bX_s)}$,
the latter equation becomes
\be
  {\ddot \psi}_R (t,\bX_1,\bX_2,..., \bX_r) 
  + \sum_{m=1}^r \sum_{n=1}^r \om_{\bx_m} \om_{\bx_n} {\psi}_R (t,\bX_1,\bX_2,..., \bX_r) = 0 .
\lbl{B9}
\ee

Introducing now the notation
$\phi(t,\bX_1,\bX_2,..., \bX_r) \equiv \psi_R (t,\bX_1,\bX_2,..., \bX_r) \equiv {\tl \vphi} (t_1,t_2,...,t_r,\bX_1,\bX_2,..., \bX_r)$
and using  the relation (\ref{3.22a}), which implies
\be
  \frac{\dd^2 \phi}{\dd t^2} = \sum_{m,n} \frac{\p^2 {\tl \vphi}}{\p t_m \p t_n} ,
\lbl{B10}
\ee
Eq.\,(\ref{B9}) can be written in the form
\be
  \sum_{M,n} \left ( \frac{\p^2 {\tl \vphi}}{\p t_m \p t_n} + \om_{\bx_m} \om_{\bx_n} {\tl \vphi} \right ) = 0 .
\lbl{B11}
\ee
In the latter equation is embraced the multiparticle Klein-Gordon equation (\ref{3.11})
\be
  \frac{\p^2 {\tl \vphi}}{\p t_n^2} + \om_{\bx_n}^2 {\tl \vphi} = 0 .
\lbl{B12}
\ee

For illustration let us now consider a flat space solution of the above equation,
$$
  {\tl \vphi}(t_1,...,t_r,\bx_1,...,\bx_r) \hs{9cm}$$
\be  
 \hs{2.5cm}  = \int \dd^3 \bp_1 ... \dd^3 \bp_r \left ( c(\bp_1,...,\bp_r)
  {\rm e}^{-\sum_k p_k x_k} + c^*(\bp_1,...,\bp_r){\rm e}^{\sum_k p_k x_k} \right ) .
\lbl{B13}
\ee
Taking succesively the derivative with respect to $t_n$ and $t_m$, we obtain
$$
   \frac{\p^2 {\tl \vphi}}{\p t_m \p t_n} =  \int \dd^3 \bp_1 ... \dd^3 \bp_r (- \om_{\bp_m} \om_{\bp_n})\left ( c(\bp_1,...,\bp_r)
  {\rm e}^{-\sum_k p_k x_k} \right . \hs{3cm}$$
\be  
 \hs{7cm}  \left . + \,  c^*(\bp_1,...,\bp_r){\rm e}^{\sum_k p_k x_k} \right ) .
\nonumber
\ee
\be
   = - \om_{\bx_m} \om_{\bx_n} {\tl \vphi} ,\hs{7cm}
\lbl{B14}
\ee
where $\om_{\bp_n} = \sqrt{m^2 + \bp_n^2}$, Summing over $m,n$ we obtain precisely Eq.(\ref{B11})
(apart from the fact that the above illustration is done for a flat space solution,
while Eq.\,(\ref{B11}) holds in curved space as well).

\end{document}